\documentclass[a4paper,twocolumn,11pt,accepted=0000-00-00]{quantumarticle}
\pdfoutput=1
\usepackage[utf8]{inputenc}
\usepackage[english]{babel}
\usepackage[T1]{fontenc}
\usepackage{hyperref}
\usepackage{amsfonts}
\usepackage{amssymb}
\usepackage{mathrsfs}
\usepackage[intlimits]{amsmath}
\usepackage{bm, bbm}
\usepackage{tikz}
\usepackage{lipsum}
\usepackage[numbers]{natbib}

\def\dt{{\rm d}\,}

\newcommand{\ket}[1]{| #1 \rangle}
\newcommand{\bra}[1]{\langle #1 |}

\def\duzomniejsze{<\kern-.7mm<}
\def\duzowieksze{>\kern-.7mm>}

\def\textbf#1{{\bf #1}}
\def\be{\begin{equation}}
\def\ee{\end{equation}}
\def\eea{\end{array}}
\def\bea{\begin{array}}
\newcommand{\bei}{\begin{itemize}}
	\newcommand{\eei}{\end{itemize}}
\newcommand{\bee}{\begin{enumerate}}
	\newcommand{\eee}{\end{enumerate}}

\def\1{\openone}

\def\tr{{\rm Tr}}

\def\>{\rangle}
\def\<{\langle}

\def\dt#1{{{\kern -.0mm\rm d}}#1\,}
\def\squareforqed{\hbox{\rlap{$\sqcap$}$\sqcup$}}
\def\qed{\ifmmode\squareforqed\else{\unskip\nobreak\hfil
		\penalty50\hskip1em\null\nobreak\hfil\squareforqed
		\parfillskip=0pt\finalhyphendemerits=0\endgraf}\fi}

\newtheorem{lemma}{Lemma}
\newtheorem{theorem}[lemma]{Theorem}
\newtheorem{main result}[lemma]{Main result}
\newtheorem{proposition}[lemma]{Proposition}
\newtheorem{definition}{Definition}

\newtheorem{fact}[lemma]{Fact}

\def\bep{\begin{proposition}}
	\def\eep{\end{proposition}}
\def\bel{\begin{lemma}}
	\def\eel{\end{lemma}}

\def\bet{\begin{theorem}}
	\def\eet{\end{theorem}}
\def\bed{\begin{definition}}
	\def\eed{\end{definition}}
\def\bef{\begin{fact}}
	\def\eef{\end{fact}}

\begin{document}

\title{Purifying teleportation}

\author{K. Roszak}
\affiliation{Institute of Physics (FZU), Czech Academy of Sciences, Na Slovance 2, 182 00 Prague, Czech Republic}
\orcid{0000-0002-9955-4331}
\email{roszak@fzu.cz}
\author{J. K.~Korbicz}
\affiliation{Center for Theoretical Physics, Polish Academy of Sciences, Aleja Lotnik{\'o}w
	32/46, 02-668 Warsaw, Poland}
\orcid{0000-0003-2084-7906}
\email{jkorbicz@cft.edu.pl}
\maketitle

\begin{abstract}
Coupling to the environment typically suppresses quantum properties of physical systems via decoherence mechanisms. This is one of the main obstacles in practical implementations of quantum protocols. In this work we show how decoherence effects can be reversed/suppressed during quantum teleportation in  a network scenario. Treating the environment quantumly, we show that under a general pure dephasing coupling, performing a second teleportation step can probabilistically reverse the decoherence effects if certain commutativity conditions hold. This  effect is purely quantum and most pronounced for qubit systems, where in 25$\%$  of instances   
the decoherence can be reversed completely. As an example, we show the effect in a physical model of a qubit register coupled to a bosonic bath. We also analyze general $d$-dimensional systems, identifying all instances of decoherence suppression. Our results are  proof-of-concept but we believe will be relevant for the emerging field  of quantum networks as teleportation is the key building block of network protocols.
\end{abstract}

\section{Introduction}
The role of the environment and decoherence in quantum systems has been a subject of intensive studies due to its great importance for both fundamental understanding of the quantum theory \cite{joos03,schlosshauer07,zurek09} and  for practical implementations of various landmark quantum effects \cite{landig18,burnett19,schlor19,liu21,chen22}, which we believe will lead to novel technological developments.  One of such landmark effects is the widely-known quantum state teleportation.  It relies on quantum entanglement as a resource for transmitting an unknown quantum state from one particle to another via a well defined protocol. Since its discovery \cite{bennett93}, quantum teleportation has been extensively studied theoretically, e.g. in \cite{bowen01,verstraete03,bandyopadhyay06,prakash12,zhang07a}, and demonstrated experimentally \cite{ren17,liu20,llewellyn20,hu20,fiaschi21,langenfeld21}.  It has become especially important recently due to its fundamental role for the emerging quantum networks \cite{cacciapuoti20,hermans22}. One of the real-life challenges in the practical implementation of teleportation is the inevitable coupling to the environment and the resulting decoherence. The entangled resource, necessary for establishing the teleportation channel, is especially sensitive to decoherence  because entanglement requires some level of non-locality, i.~e.~coherences between distinct states of two completely distinguishable particles, and non-local phase relations are always more fragile than local ones, since they are easier to distinguish for the environment. This motivated more realistic studies of quantum teleportation with noisy, non-ideal entanglement resources
\cite{bowen01,verstraete03,bandyopadhyay06,prakash12}. The role of the environment has been analyzed e.g. in \cite{verstraete03,bandyopadhyay06}, showing, as one would expect, its detrimental influence on the teleportation fidelity. However, this does not always have to be so and 
we can take advantage of the quantum nature of the interactions to suppress decoherence.

In this work we show how to reverse  detrimental effects of coupling to the environment during teleportation. In particular, we demonstrate that repeating a noisy teleportation, i.e. performing  a second non-ideal teleportation process coupled to the environment, can, under certain general conditions, probabilistically reverse the decoherence effects instead of accumulating them. This counterintuitive effect is purely quantum and based on a fully quantum treatment of the environment. In particular, we assume a realistic coupling via a pure dephasing interaction \cite{kawakami16,malinowski17,yulin18,touzard19}, which appears in many physical systems, e.g. a spin coupled to a bosonic bath.  The effect is most pronounced for the lowest dimensional quantum systems, i.e. spin-$1/2$ or qubits, where we show that in 25$\%$ of the instances, the second teleportation completely purifies the teleported state, resulting in a perfect copy of the input state. For higher dimensional systems, the effect is still there but less pronounced.

This decoherence reversal effect has been unnoticed before since environments' influence has been usually modeled with the help of quantum channels or other methods operating only at the level of the reduced density matrix of the system of interest \cite{bowen01,verstraete03,bandyopadhyay06,prakash12}. The problem with this treatment is that it  effectively reduces the environment to a source of noise, which as turns out is not always sufficient for a description of its effects. A fully quantum inclusion of the environment, on the other hand, has lead to such experimentally important decoherence reversal techniques as the spin echo (see e.g. \cite{vandersypen05,shi18,gardner20,roszak21}) and dynamical decoupling \cite{pokharel18,kennedy20,ma21,hahn22}. In the case of teleportation, purely quantum treatment of the environment was analyzed in \cite{harlender22}, where the teleportation of correlations with the environment in a cyclic protocol where studied, but the reversal effects were not noticed.

\section{Purifying teleportation protocol.}

\begin{figure}[t]
	\centering
	\includegraphics[width=0.45\textwidth]{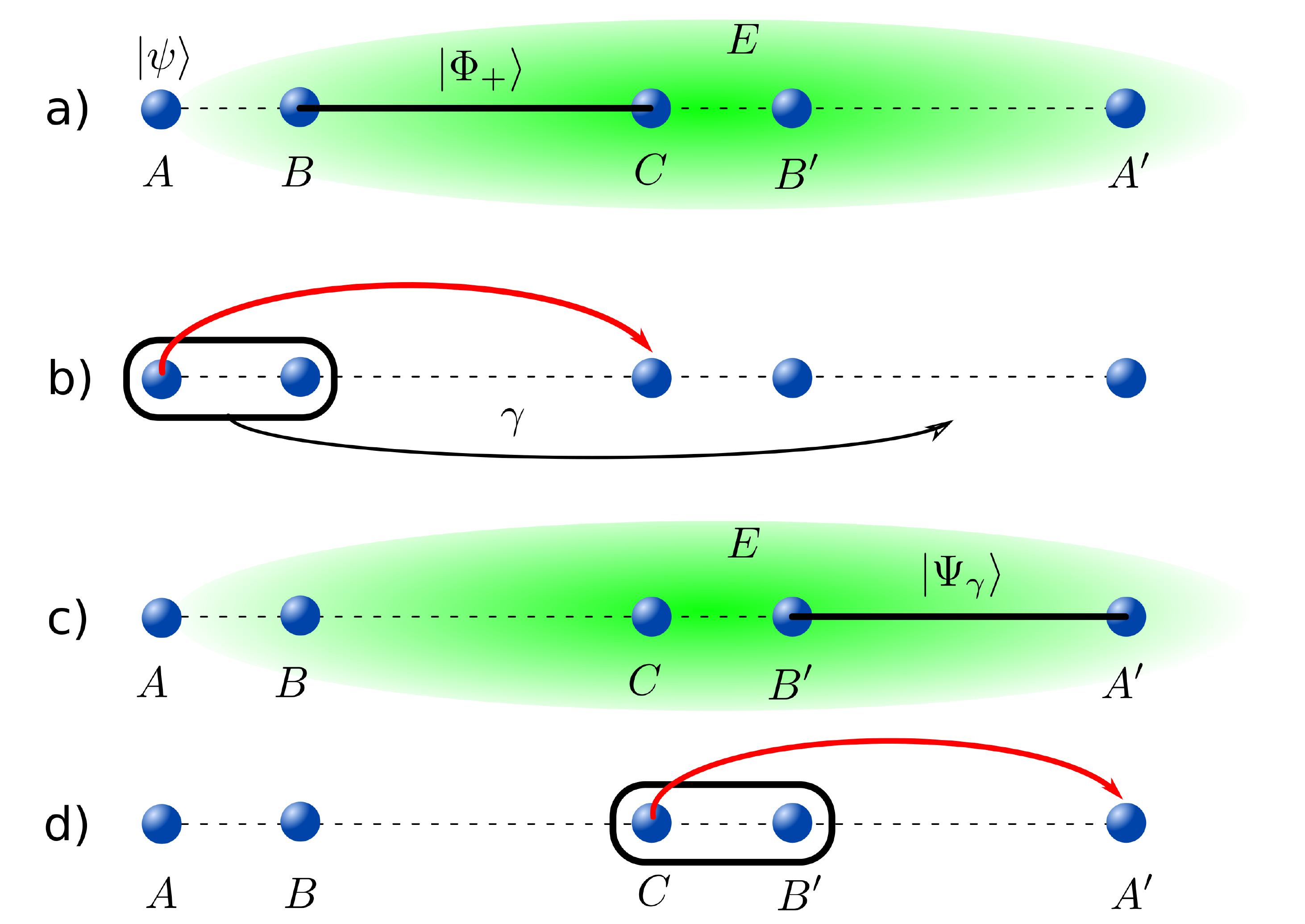}
	\caption{The two-step teleportation protocol along a chain allowing for purification. a) Initialization and the action of decoherence on the entangled resource $BC$. The state to be teleported is $\ket\psi_A$. b) Bell measurement on qubits $AB$, followed by the teleportation $A \to C$ and a feed-forward of the Bell measurement outcome $\gamma$ to qubits $A'B'$. c) Initialization of qubits $A'B'$ in the Bell state $\ket{\Psi_\gamma}$, corresponding to the measurement outcome in the previous step, followed by the action of decoherence on $A'B'$. d) The second, final teleportation 
		$C \to A'$.}
	\label{fig1}
\end{figure}

We first consider a spin-$1/2$ system, composed of three particles (qubits): $A$  in an unknown state $\ket\psi$ to be teleported and $BC$ prepared initially in a maximally entangled state (the entanglement resource). The environment $E$ is included in the quantum description and we initialize the whole system in the state:
\be \label{stan0}
\sigma(0)=\ket{\psi}_A\bra\psi\otimes\ket{\Phi_+}_{BC}\bra{\Phi_+}\otimes R(0),
\ee
where 
\be \label{stan_t}
\ket{\psi}_A=\alpha\ket 0+\beta\ket 1
\ee
is the state to be teleported, $\ket{\Phi_+}\equiv 1/\sqrt 2(\ket{00}+\ket{11})$ is one of the Bell states in some chosen basis $\ket 0$, $\ket 1$, and $R(0)$ is an arbitrary state of the environment.
The particular choice of the Bell state is irrelevant for the idea and we have chosen $\ket{\Phi_+}$ only for definiteness' sake (the remaining choice of
$\ket{\Psi_\pm}$ is analyzed in Appendix \ref{Psi}). 
The interaction with the environment is modeled via a general pure decoherence (dephasing) dynamics:
\be\label{H}
H_{dec}=\sum_{i,j=0,1} \ket{ij}\bra{ij} \otimes V_{ij},
\ee
where $V_{ij}=V_{ij}^\dagger$ are some observables on the environment $E$. The resulting evolution is controlled by the two-qubit state:
\be \label{U}
U(\tau)=\sum_{i,j=0,1} \ket{ij}\bra{ij}\otimes w_{ij}(\tau), \quad w_{ij}(\tau)\equiv e^{-i \tau V_{ij}}.
\ee
This is a very general formalism that can describe any source of pure dephasing: entangling or separable \cite{roszak15}, with or without a self Hamiltonian of the environment in $V_{ij}$, systems interacting with the same or separate environments, Markovian and non-Markovian etc. 

The two-step noisy teleportation procedure \cite{harlender22}  is constructed as follows; see Fig.~\ref{fig1}. The decoherence process \eqref{U} acts for some time $\tau$ on qubits $BC$, coupling them with the environment. As mentioned before, we assume that the decoherence affects only the 
entangled resource as it is more sensitive than a single system (see e.g. \cite{reina02}). Afterwards, following the standard teleportation procedure, a two qubit Bell measurement 
is performed on $AB$ and depending on the measurement outcome, the appropriate corrective unitary 
operation is applied to the qubit $C$. This concludes the first step. The outcome, conditioned on the Bell measurement result, is:
\be \label{stan1}
\sigma_\gamma(\tau)=\ket{\Psi_\gamma}_{AB}\bra{\Psi_\gamma}\otimes\rho^{(\gamma)}_{CE}(\tau).
\ee
Here $\ket{\Psi_\gamma}_{AB}$ denotes the Bell state that was measured. The state of $CE$ depends on the measurement result $\gamma$ as the corrective unitary operations are applied to $C$ only and leave the environment part alone. This leads to an effective outcome-dependent permutation on the $E$ side:
\be
\rho^{(\gamma)}_{CE}(\tau)=\sum_{j,j'=0,1}\psi_j\psi_{j'}^*\ket j\bra{j'}\otimes R_{j\oplus m, j'\oplus m}\label{stanCE}
\ee
where $\psi_k$ are from eq.~\eqref{stan_t}, $R_{ij}=w_{ii}(\tau) R(0) w_{jj}^\dagger(\tau)$ are matrices on the environment (which in principle can be very large), $\oplus$ denotes the addition modulo 2, and $m=0$ for $\ket{\Phi_{\pm}}$ outcomes and $m=1$ for $\ket{\Psi_{\pm}}$.
Obviously without decoherence $\rho_{CE}(\tau)$ would be a product state 
$\rho_{CE}(\tau)=\ket{\psi}_{C}\bra{\psi}\otimes R(0)$, but in the presence of 
decoherence \eqref{U}, the state \eqref{stanCE} gets correlated, possibly entangled \cite{roszak15,roszak18}.
This, as expected, alters the state of the qubit $C$, which for $\gamma=\Phi_{\pm}$ reads:
\be \label{state_1}
\varrho_C(\tau)=\begin{pmatrix} |\alpha|^2 & \alpha\beta^* c(\tau) \\
	\alpha^*\beta c^*(\tau) & |\beta|^2 \end{pmatrix},
\ee
where
\be\label{c} 
c(\tau)\equiv \tr_E [w_{11}^\dagger(\tau) w_{00}(\tau) R(0)]
\ee
is a decoherence factor between the states $\ket{00}$, $\ket{11}$ (for $\gamma=\Psi_{\pm}$,
factors $c(\tau)$ and $c^*(\tau)$ are interchanged; see Appendix \ref{allstates} for details).
Decoherence clearly affects the fidelity $F_1 =\bra\psi \varrho_C\ket \psi$ between the teleported state \eqref{state_1} and the original state 
\eqref{stan_t}:
\begin{equation}\label{F1}
F_1(\tau)=1-2|\alpha\beta|^2\left[1-\mathrm{Re}\left[c(\tau)\right]\right].
\end{equation}
We note that the ideal decoherence, $c(\tau)=0$, leads to $F_{min}=1-2|\alpha\beta|^2$, 
which is in general greater than zero as we are only destroying the coherences
in \eqref{state_1} and that  is not enough to make \eqref{state_1} orthogonal to \eqref{stan_t}. 

We now perform the second noisy teleportation. We prepare the additional qubits $A'B'$ in the Bell state corresponding to the first Bell measurement outcome and this state will serve as the entanglement resource for the teleportation. Thus the initial state for the second step is of the form \eqref{stan1} but with the resource prepared on a new system to propagate the state further rather than coming back to the original qubit $A$ like it was in \cite{harlender22}: 
\be \label{stan1'}
\sigma'_\gamma(\tau)=\ket{\Psi_\gamma}_{A'B'}\bra{\Psi_\gamma}\otimes\rho^{(\gamma)}_{CE}(\tau).
\ee
The decoherence process \eqref{U} is assumed to couple to the same environment and last for the same time $\tau$ as in the first teleportation and this is crucial for our protocol.  This is of course an idealized situation, but we are concerned with a proof-of-principle here, investigating how much suppression is theoretically possible. Coupling to the same environment can be a good approximation to the situations when the nodes are close enough compared to the effective range of the interaction with the environment. Robustness with respect to small time mismatches will be discussed below. We assume that decoherence again affects the entangled resource $A'B'$ as the most sensitive. After measuring qubits $B'C$ in the Bell basis, we apply a correction unitary to $A'$ but this time the unitaries depend not only on the result of the second Bell measurement, but also on the first one as the latter determines the entanglement resource in \eqref{stan1}; see Appendix \ref{allstates}. The final state of $A'$ depends now on the measurement outcomes in both steps but is the same for either sign of $\ket{\Phi_{\pm}}_{AB}$ 
and of $\ket{\Psi_{\pm}}_{AB}$ in \eqref{stan1}, reducing the number of all possibilities to four.
For $\ket{\Phi_{\pm}}_{AB}$ in \eqref{stan1}, the second decoherence process will again be
governed by the $w_{00}(\tau), w_{11}(\tau)$  and the state of $A'$
will be given by \eqref{state_1} but with $c(\tau)$ replaced by 
(derived in the Appendix \ref{allstates}):
\begin{align}
C_{\Phi}(\tau)&\equiv\tr_E[ w_{11}^{\dagger 2}(\tau) w_{00}^2(\tau) R(0)],\label{C1}\\
C_{\Psi}(\tau) &\equiv \tr_E[ w_{11}^\dagger(\tau) w_{00}^\dagger(\tau) w_{11}(\tau) w_{00}(\tau)R(0)],\label{C2}
\end{align}
depending on the result of the second Bell measurement. $C_{\Phi}, C_{\Psi}$ correspond, respectively, to $\ket{\Phi_{\pm}}$ and $\ket{\Psi_{\pm}}$.	An interesting effect happens for
\be \label{com}
[w_{11}(\tau), w_{00}(\tau)]=0,									  
\ee							  
when
\be 
C_{\Psi}(\tau)=\tr_E R(0) =1 \label{PT}
\ee
strictly and irrespectively of any other parameters.
Thus, for those instances when $\ket{\Psi_{\pm}}$ is measured, the second noisy teleportation actually purifies the  state \eqref{state_1} back to the initial form instead of adding more noise as it is the case for $\ket{\Phi_{\pm}}$ measurements.

This does not happen for outcomes $\ket{\Psi_\gamma}_{AB}=\ket{\Psi_{\pm}}_{AB}$ of the first measurements as then the decoherence factors read:
\begin{align}
C'_{\Phi}(\tau)&\equiv\tr_E[w_{00}^\dagger(\tau)w_{01}^\dagger(\tau)w_{10}(\tau) w_{11}(\tau)R(0)],
\label{C1'}\\
C'_{\Psi}(\tau)&\equiv \tr_E[ w_{00}^\dagger(\tau)w_{10}^\dagger(\tau)  w_{01}(\tau) w_{11}(\tau)R(0)],\label{C2'}
\end{align}
and no commutativity condition can make it equal to unity for all $\tau$. It is easy to check that the probabilities of each of the four possible outcomes \eqref{C1}, \eqref{C2}, \eqref{C1'}, \eqref{C2'}  are all equal to $1/4$.
Thus, the purifying teleportation \eqref{PT} happens in $25\%$ of the cases. The the statistical state at the end of our protocol $\varrho_{A'f}(\tau)$ is the mixture with equal weights $1/4$ of the conditional states corresponding to the each set of the measurement outcomes (see Appendix \ref{allstates}). Its fidelity with respect to the original state \eqref{stan_t} is given by the average:
\begin{equation}\label{F2}
F_2(\tau)=1-2|\alpha\beta|^2\left[1-\mathrm{Re}\left[C_{av}(\tau)\right]\right],
\end{equation}
where
\begin{equation}
\label{cav2}
C_{av}(\tau)\equiv\frac{1}{4}\left[1+C_\Phi(\tau)+C'_{\Phi}(\tau)+C'_{\Psi}(\tau)
\right].
\end{equation}
In many interesting situations, like the one described below, all the $C(\tau)$-factors in \eqref{F1} and \eqref{cav2} tend asymptotically to zero with $\tau$. Then the presence of the constant factor $1/4$ in \eqref{cav2} causes  $F_2 > F_{min}$ asymptotically and as a result $F_2> F_1$.  Thus the purifying effect reveals itself not only in postselected events but also in the average state.

It is clear how the purifying effect propagates down a network: After a third teleportation there will be instances when the decoherence is reduced to single-step values, after a fourth teleportation there will again appear complete reversals, and so on.

The commutativity condition \eqref{com} is not as restrictive as it may seem. Obviously it is satisfied when the environmental observables in \eqref{H} commute: $[V_{ij},V_{kl}]=0$, which includes such a fundamentally important case \cite{schlosshauer07} as $H\approx A\otimes V$, where $A$, $V$ are some observables on the system and the environment respectively. Let us also briefly address the question of robustness, leaving more detailed analysis to a future work. First of all, decoherence factors $C_\Psi(\tau)$, $C_\Phi(\tau)$ are continuous functions of $V_{ij}$, $R(0)$, and $\tau$ and thus small variations in these parameters  will lead to small variations of the coefficients. As an example let us examine a small time mismatch. Assuming for simplicity $[V_{00},V_{11}]=0$, which ensures \eqref{com} for all times, we obtain:
\begin{align}
&&|C_{\Psi}(\tau_1,\tau_2)|^2=\left|\tr_E\left[w_{11}(\Delta\tau) w_{00}(-\Delta\tau)R(0)\right]\right|^2\nonumber\\
&&\approx 1-\Delta\tau^2\langle\Delta(V_{11}-V_{00})^2\rangle \approx e^{-\Delta\tau^2 \langle\Delta(V_{11}-V_{00})^2\rangle},\nonumber\\
\end{align}
so that to the lowest order, the departure from the ideal canceling \eqref{PT} decays quadratically with $\Delta \tau=\tau_2-\tau_1$ and with the rate controlled by the variance of $(V_{11}-V_{00})$ in the initial state $R(0)$.

\begin{figure}[t]
	\centering
	\includegraphics[width=0.5\textwidth]{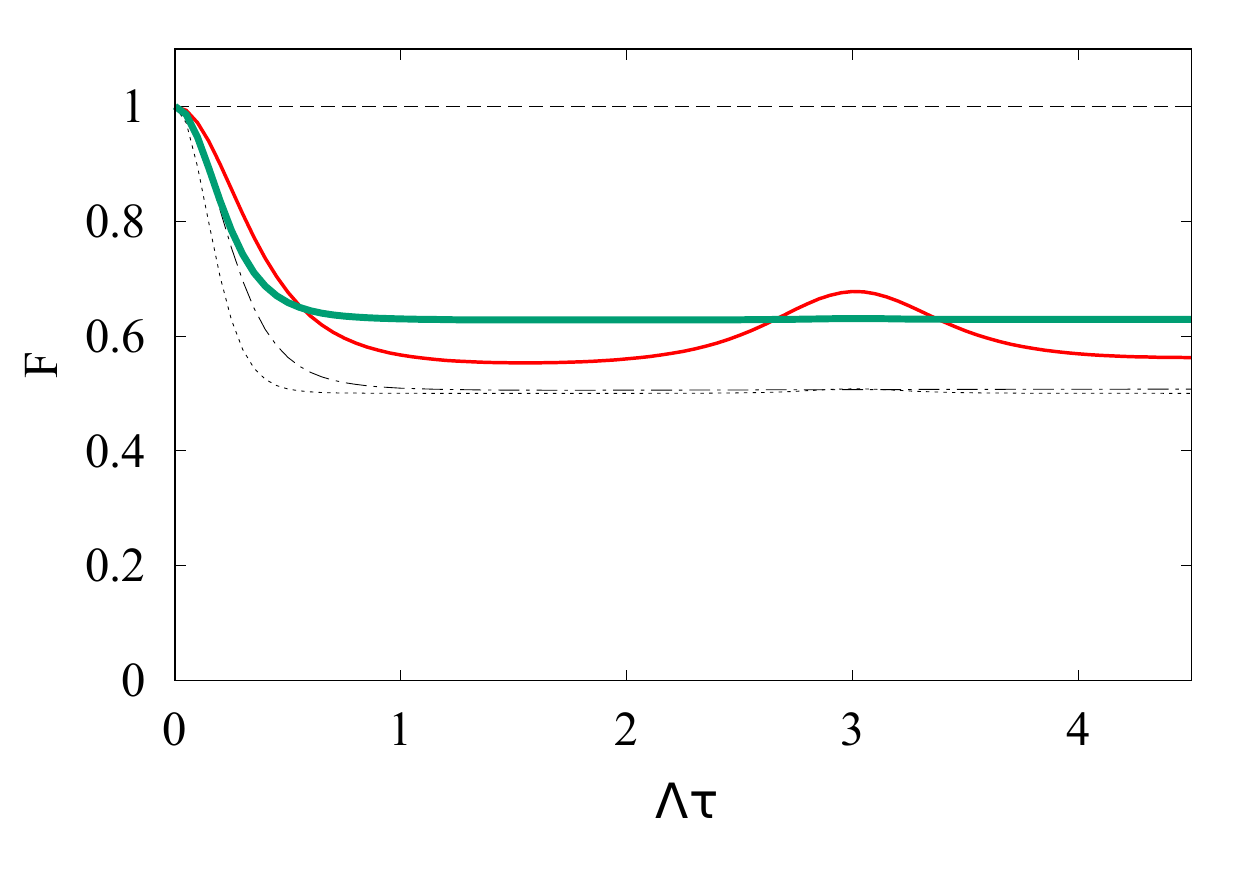}
	\caption{[Color online] Example of purifying teleportation in the spin-boson model. Red solid line shows  the single-step fidelity \eqref{F1} and green, thick solid line the two-step average fidelity \eqref{F2} as functions of  decoherence duration $\tau$. Time $\tau$  is plotted in cutoff units $\Lambda^{-1}$, temperature is set to $T=1/10\Lambda$, time-of-flight $\bar t=3\Lambda^{-1}$, and Ohmicity parameter is $s=3$. Initial state \eqref{stan_t} is chosen with $\alpha=\beta=1/\sqrt{2}$. The plot of \eqref{F1} shows a well known recoherence impulse at $\tau=\bar t$ but asymptotically it decays to a value lower than the assymptotic of  \eqref{F2}. Thus, for long enough duration $\tau$ of the decoherence processes, the second application of decoherence $+$ teleportation actually increases the fidelity of the output state w.r.t.~the one-step process.  To better visualize how it happens, we plot all the components of \eqref{F2}, each entering with the equal weight of $1/4$: Black dotted line corresponds to the partial fidelity associated with $C_\Phi(\tau)$, dashed-dotted line to both factors \eqref{C'}, and the dashed line to the constant term \eqref{PT}. Both $C_\Phi(\tau)$- and $C'_{\Phi,\Psi}$-dependent factors decay quickly to $F_{min}=1/2$ as expected, but the presence of the constant term  makes final fidelity asymptoticaly higher than the first-step fidelity. For short durations $\tau\ll \Lambda^{-1}$, the situation is reversed and $F_2(\tau)<F_1(\tau)$. }
	\label{fidelity}
\end{figure}

\section{Spin register example}
We consider a well known model of a two-qubit register coupled to a bosonic bath (see e.g. \cite{reina02}), described by the Hamiltonian:
\be
H=\sum_{m=1,2}\frac{\sigma_z^{(m)}}{2}\otimes \sum_{k}(g_{mk}a_k^\dagger+g_{mk}^*a_k) + \sum_k H_k,
\ee
where $\sigma_z^{(m)}$ are the $m$-th qubit $z$-axis Pauli matrices, $a_k,a_k^\dagger$ are the $k$-th mode annihilation and creation operators, $g_{mk}$ are complex coupling constants, and  $H_k=\omega_k a_k^\dagger a_k$. 
Here $V_{ij}=\sum_k(\epsilon_i g_{1k} a_k^\dagger +\epsilon_j g_{2k}a^\dagger_k+h.c.+H_k)$, where $\epsilon_i=\mp1/2$ for $i=0,1$ respectively. Although $[V_{ij},V_{lm}]\ne 0$, one checks that condition \eqref{com}
is satisfied: $[w_{11},w_{00}]=0$ (also $[w_{01},w_{10}]=0$ as any other pair with co-linear $\bm \epsilon$, $\bm \epsilon'$).  Assuming the initial state of the 
environment to be thermal, $R(0)=1/Ze^{-\beta\sum_k H_k}$, with the inverse temperature $\beta$, the decoherence factor \eqref{c} has been found in e.g. \cite{reina02,tuziemski19}:
\begin{align}
& \ln c(\tau)=\nonumber\\
&-2\int_0^\infty d\omega J(\omega) \frac{1-\cos\omega\tau}{\omega^2}\left(1+\cos\omega\bar t\right)\coth\frac{\omega\beta}{2},  \label{c-model} 
\end{align}
Here $\bar t \equiv k|r_1-r_2|/\omega$ is the time-of-flight between the qubit positions, assuming that ${g}_{mk}=g_ke^{-kr_m}$, which is the case e.g. for the electromagnetic bath in the dipole approximation. Further assuming the
spectral density of the form $J(\omega)=\omega(\omega/\Lambda)^{s-1}e^{-\omega/\Lambda}$, where $\Lambda$ is the cut-off frequency, the above integral can be calculated analytically \cite{tuziemski19}, but we will not need it here.
The factors $C_\Phi, C'_\Phi, C'_\Psi$ can be also calculated 
(see Appendix \ref{Cfactors}). 
Like $c(\tau)$, they are real and read: $\ln C_\Phi(\tau) =4 \ln c(\tau)$,
\begin{align}
\label{C'}
\ln C'_\Phi(\tau)&=\ln C'_\Psi(\tau) \\
&= -4\int d\omega J(\omega) \frac{1-\cos\omega\tau}{\omega^2}\coth\frac{\omega\beta}{2}. \nonumber  
\end{align}
Note the absence of the time-of-flight $\bar t$ above, indicating collective decoherence, when the register behaves like a single system \cite{reina02}. Thus, after the second step there are three types of decoherence present: No decoherence \eqref{PT}, true $2$-qubit decoherence \eqref{c-model}, and collective decoherence \eqref{C'}. Sample plots of the fidelities, corresponding to those three processes, as well as the average fidelity \eqref{F2} and the fidelity after the first teleportation \eqref{F1}, are presented in Fig.~\ref{fidelity}. We see that for short durations $\tau\ll \Lambda^{-1}$ of the decoherence processes, teleportation fidelity after one step is higher than after two steps as one would expect. However, for a longer duration
all the contributions to $F_2(\tau)$ from $C_\Phi, C'_\Phi, C'_\Psi$  decay quickly to $F_{min}$, but the constant term due to \eqref{PT} forces that $F_2 >F_1$ after the initial transients, including a propagation of an disturbance between the qubits. 
Thus, in this example, the purifying effect of the second teleportation is present in the statistical state too, if the decoherence duration is long enough.

\section{Generalization to dimension $d$}
The generalized teleportation protocol is defined by the following $d^2$ measurement vectors and correction unitaries \cite{bennett93}:
\begin{align}
\ket{\Psi_{nm}}&=\frac{1}{\sqrt d}\sum_{j=0}^{d-1} e^{\frac{2\pi ijn}{d}} \ket{j,j\oplus m}, \label{Pnm_main}\\
U_{nm}&=\sum_{j=0}^{d-1} e^{\frac{2\pi ijn}{d}} \ket j \bra{j\oplus m}, \label{Unm_main}
\end{align}
where $n,m\in [0,d-1]$, same as the basis indicies $j,j',k,k',\cdots$, and $\oplus$ is the addition modulo $d$. We assume the following initial state:
\be \label{stan0d_main}
\sigma(0)=\ket{\psi}_A\bra\psi\otimes\ket{\Psi_{00}}_{BC}\bra{\Psi_{00}}\otimes R(0),
\ee
where $\ket\psi=\sum_{j=0}^{d-1}\psi_j\ket j$ and $\ket{\Psi_{00}}=1/\sqrt d \sum_j \ket{jj}$. The final state after the two-step teleportation reads 
(see Appendix \ref{ddit}):
\be
\varrho^{A'}_{mm'}=\sum_{jj'} \psi_j\psi^*_{j'}  C^{mm'}_{jj'} \ket j \bra{j'}, \label{rhofinal_main}
\ee
where:

\begin{align}
C^{mm'}_{jj'}(\tau)=&
\tr_E\left[w_{j\oplus M,j\oplus M\oplus m}w_{j\oplus m,j\oplus m }R(0)\right.
\nonumber\\
& \times\left. w^\dagger_{j'\oplus m, j'\oplus m}w^\dagger_{j'\oplus M,j'\oplus M\oplus m}\right]\label{bigC_main}
\end{align}
are the decoherence factors between the levels $jj'$, with $m,m'$ labeling the results of the first and the second Bell measurements (the states are independent of the $n$-indicies), and  $M=\left[d-(m\oplus m')\right]\text{mod}\, d$. A detailed analysis
found in the Appendix \ref{ddit} shows that under the commutation condition $[w_{ii},w_{jj}]=0$, the decoherence suppression/reduction happens only when $m=0$ and has the following pattern:
i) For every $m'$, there will be  $d$ levels $jj'$ in the corresponding state $\varrho_{0m'}$ \eqref{rhofinal_main}, such that the decoherence factors \eqref{bigC_main} between them reduce to the one-step form, i.e. like if only one decoherence process has happened; ii) Additionally, when $d$ is even, the state corresponding to $m'=d/2$ will have $d/2$ of its coherences  fully restored to the original form like if no decoherence has happened at all. These findings have two implications: a) using higher-$d$ systems gives no advantage in terms of noise cancellation/suppression  in the network scenario considered here; to the contrary, in relatively less instances coherence losses can be suppressed/reversed than for qubits;   b) if higher-$d$ systems must be used for some reason, we provide the list of instances when the noise suppression happens. 

\section{Concluding remarks}
We have presented a decoherence reversal/suppression mechanism in the quantum teleportation protocol, which allows to use quantum nature of decoherence to our advantage. Working in a quantum open systems framework, we have studied a fully quantum teleportation scheme, where the environment is modeled as a quantum system, and shown that a  repetition of the teleportation protocol can, under suitable conditions, reverse the detrimental decoherence effects, or at least suppress them, instead of accumulating. This rather counterintuitive effect is due to the quantumness of the environment and is probabilistic. That is, it is guaranteed to occur when a certain measurement result is obtained, in the same manner as it happens e.g. in the famous Elitzur-Vaidman interaction-free measurement \cite{elitzur93}. It is also most visible for the lowest dimensional quantum systems, which reflects its highly non-classical nature: For qubits there is a complete decoherence reversal with probability $1/4$  while for quitrits in $1/3$ of the instances the decohrence reduces to the single-step value.  

Let us briefly discuss a connection to other decoherence reversal and purification techniques. Spin-echo (see e.g. \cite{vandersypen05} or \cite{roszak21} for a more relevant exposition), is an  experimental technique, widely used to prolong coherence times. While there are certain similarities, e.g. both protocols rely on a two-step procedure and commutativity of the conditional environmental evolutions, there are fundamental differences. Firstly, decoherence here is acting on physically distinct systems in each step, while in the spin-echo it acts on the same system. Secondly, and most importantly, unlike the spin-echo, which is basically a single-system effect, purifying teleportation is a multipartite effect, which fully relies on entanglement. The latter allows to shift decoherence effects from one system to another \cite{harlender22} and opens the chance for their cancellation. Superbroadcasting effect \cite{dariano05}, is in turn a purification effect, where $N$ input copies of a sufficiently mixed qubit state can be broadcasted in form of correlated copies to $M>N$ receivers, such that each of the local states has greater purity, $\tr\varrho^2$, than the initial state. The difference is that superboradcasting assumes an inherent noise, which must be present at the input, and transfers it to correlations among the copies, purifying along a Bloch vector, while purifying teleportation physically reduces the noise dynamically acquired from the environment.

The significance of our effect is two-fold. First, it shows how to improve the fidelity of a single noisy teleportation -- quite surprisingly by performing a second one (this is most notable for qubits, but also for other even-dimensional systems). 
But more interestingly  it shows, at least in theory, a way to mitigate teleportation noise in quantum networks, where teleportation is the key primitive for quantum information distribution. By a proper design of the teleportation processes, the noise can be reduced instead of accumulated during multiple teleportations between the nodes. Of course the practical issues must be first addressed.  We have already briefly commented on the time mismatch.  Another important  point  is the common environment. For the effect to happen there must be at least some overlap between the environments in both steps and we expect  the magnitude of the effect will approach the theoretical bounds presented here with the growing overlap. The common environment may be also viewed as a condition on the physical node separation, i.e. that it is not too big compared to the effective interaction range, as we have already mentioned. 
Other practical issues include  possible difference in the coupling observables and the neglected decoherence of the individual particles. We leave those important questions for a future work.

{\it Acknowledgements}---JKK acknowledges the financial support of the Polish National Science Center (NCN) through the grant no.
2019/35/B/ST2/01896 and the hospitality of Institute of Physics (FZU) of the Czech Academy of Sciences. KR acknowledges support from project 20-16577S of
the Czech Science Foundation.

\onecolumn
\appendix

\section{The two-step teleportation procedure} \label{allstates}

We present the details of the two-step teleportation; see Fig.~\ref{fig1}. We initialize the whole system in the state: 
\be 
\sigma(0)=\ket{\psi}_A\bra\psi\otimes\ket{\Phi_+}_{BC}\bra{\Phi_+}\otimes R(0),
\ee
and then apply the decoherence process \eqref{U} to $BCE$.

Then the standard teleportation procedure is performed: A joint measurement in the Bell basis $\ket{\Phi_\pm},\ket{\Psi_\pm}$ on the qubits $AB$
followed by a corrective unitary operation of qubit $C$, conditioned on the measurement 
outcome. For completeness we recall this by now famous set of unitaries:
\be
U_{\Phi_+}=\bm 1, \ U_{\Phi_-}=\sigma_z, \ U_{\Psi_+}=\sigma_x, \ U_{\Psi_-}=i\sigma_y. \label{corr}
\ee
Consequently, the post-teleportation states of the whole system 
(each obtained with probability $1/4$) are given by: 
\begin{align}
\sigma_{\Phi}(\tau)&=\ket{\Phi_{\pm}}\bra{\Phi_{\pm}}\otimes
\left(
\begin{array}{cc}
	|\alpha|^2 R_{00,00}&\alpha\beta^* R_{00,11}\\
	\alpha^*\beta R_{11,00}&|\beta|^2 R_{11,11}
\end{array}
\right)_{CE}, \label{sigmaphi}
\\
\sigma_{\Psi}(\tau)&=\ket{\Psi_{\pm}}\bra{\Psi_{\pm}}\otimes
\left(
\begin{array}{cc}
	|\alpha|^2 R_{11,11}&\alpha\beta^* R_{11,00}\\
	\alpha^*\beta R_{00,11}&|\beta|^2 R_{00,00}
\end{array}
\right)_{CE}.\label{sigmapsi}
\end{align}
Here 
\begin{equation}
R_{ii,jj}(\tau)=w_{ii}(\tau)R(0)w_{jj}^{\dagger}(\tau)
\end{equation}
(in this Section we use a full index notation $R_{ii,jj}$ for clarity rather than the abbreviated one $R_{ij}$ used elsewhere). In the main text these states are written in a compact way in eqs \eqref{stan1}
and \eqref{stanCE}.
Note that the difference between the two states in the $CE$ part lies in the environmental
operators $R_{ii,jj}$ having their indices interchanged. This leads to the similar states of $C$ of the form \eqref{state_1} with eventually $c(\tau) \leftrightarrow c^*(\tau)$. 
Thus, the degree of coherence of qubit $C$ is the same, but there may be a difference in the phases.

We are now ready for the second teleportation process. Its aim is to propagate the state of $C$ further on the same time increasing its fidelity w.r.t.~the original state.
To this aim, we prepare additional qubits $A'B'$ in the Bell state corresponding to the first Bell measurement result. This will serve as the entanglement resource. 
Thus, the second teleportation step starts with the state \eqref{sigmaphi} or \eqref{sigmapsi} as the initial state, but with the Bell states prepared on additional qubits $A'B'$.
In both cases, the first the decoherence process acts  on the Bell states. We assume it acts for the same duration $\tau$ as in the first step. 
Then a Bell measurement is performed on qubits $B'C$, followed by the appropriate
unitary operation on qubit $A'$. Since now the initial resource for the teleportation is an arbitrary Bell state, the corrective  unitary operations will depend on the results of both measurements and will be given by some permutation of the set \eqref{corr}. We will explicitly state the correct unitaries
in what follows. 

\subsection{Second step: first measurement outcome $\ket{\Phi_\pm}$}

We take \eqref{sigmaphi} as the starting point. In this case, the decoherence is described by the same conditional environmental operators $w_{ii}(\tau)$ as in the first teleportation. 
The state of the whole system after the action of the  decoherence \eqref{U} on $A'B'E$
is given by:
	\be
	\label{sigmaphiD}
	\sigma_{\Phi_{\pm}}^{dec}(\tau)=\frac{1}{2}
	\left(
	\begin{array}{cccc}
		|\alpha|^2 R_{00,00}^{00,00}&\alpha\beta^* R_{00,11}^{00,00}&\pm|\alpha|^2 R_{00,00}^{00,11}&\pm\alpha\beta^* R_{00,11}^{00,11}\\
		\alpha^*\beta R_{11,00}^{00,00}&|\beta|^2 R_{11,11}^{00,00}&\pm\alpha^*\beta R_{11,00}^{00,11}&\pm|\beta|^2 R_{11,11}^{00,11}\\	
		\pm|\alpha|^2 R_{00,00}^{11,00}&\pm\alpha\beta^* R_{00,11}^{11,00}&|\alpha|^2 R_{00,00}^{11,11}&\alpha\beta^* R_{00,11}^{11,11}\\
		\pm\alpha^*\beta R_{11,00}^{11,00}&\pm|\beta|^2 R_{11,11}^{11,00}&\alpha^*\beta R_{11,00}^{11}&|\beta|^2 R_{11,11}^{11,11}
	\end{array}
	\right).
	\ee
The rows and columns are ordered as: $\ket{000},\ket{001},\ket{110},\ket{111}$ and the trivial rows and columns have been omitted.
The environmental operators are given by: 
\be
R_{ii,jj}^{kk,qq}(\tau)=w_{kk}(\tau)w_{ii}(\tau)R(0)w_{jj}^{\dagger}(\tau)w_{qq}^{\dagger}(\tau),
\ee
the indices $i,j$ describe the decoherence process before the first teleportation, while the indices
$k,q$ correspond to the second one.

Now the Bell measurement is performed on the qubits $B'C$ and suitable correction unitaries are applied to qubit $A'$. Like in the first step, there are four cases here, arranged in two families.

{\bf The second outcome $\ket{\Phi_\pm}$.} It is easy to check that the following operations will perform the suitable corrections on $A'$ in this case:
\be 
U_{\Phi_+\Phi_+}=U_{\Phi_-\Phi_-}=\bm 1,\  U_{\Phi_+\Phi_-}=U_{\Phi_-\Phi_+}=\sigma_z,
\ee
where $U_{\Phi\Phi'}$ corresponds to the case when the first Bell measurement resulted in $\Phi$, while the second in $\Phi'$. The final state of $AE$ is in this case given by:
\be
\label{b3}
\rho_{A'E}^{\Phi\Phi}(\tau)=\left(
\begin{array}{cc}
	|\alpha|^2 R_{00,00}^{00,00}&\alpha\beta^* R_{00,11}^{00,11}\\
	\alpha^*\beta R_{11,00}^{11,00}&|\beta|^2 R_{11,11}^{11,11}
\end{array}\right).
\ee
The coherence of $A'$ is obtained by tracing out the environment from \eqref{b3}
and is given by eq.~\eqref{C1}.

{\bf The second outcome $\ket{\Psi_\pm}$.} The corrective unitaries are now given by:
\be 
U_{\Phi_+\Psi_+}=U_{\Phi_-\Psi_-}=\sigma_x,\ U_{\Phi_+\Psi_-}=U_{\Phi_-\Psi_+}=i\sigma_y,
\ee
and lead to the following state:
\be
\label{b6}
\rho_{A'E}^{\Phi\Psi}(\tau)=\left(
\begin{array}{cc}
	|\alpha|^2 R_{00,00}^{11,11}&\alpha\beta^* R_{00,11}^{11,00}\\
	\alpha^*\beta R_{11,00}^{00,11}&|\beta|^2 R_{11,11}^{00,00}
\end{array}\right).
\ee
The coherence of $A'$ obtained from the state \eqref{b6} is given by \eqref{C2}.
For commuting conditional evolution operators $w_{ii}$, it is in this case that we observe the purification.

\subsection{Second step: first measurement outcome $\ket{\Psi_\pm}$}

The starting state is now \eqref{sigmapsi}. 
There is no qualitative difference between this case and the teleportation procedure 
described in the previous section, with the exception of the decoherence process,
which is now governed by different environmental evolution operators: $w_{01}(\tau)$
and $w_{10}(\tau)$. This is because the Bell states are now superpositions of states 
from a different subspace of the two qubit  Hilbert space.

After the decoherence process, the state \eqref{sigmapsi} takes the form 
	\be
	\label{sigmapsiD}
	\sigma_{\Psi_{\pm}}^{dec}(\tau)=\frac{1}{2}
	\left(
	\begin{array}{cccc}
		|\alpha|^2 R_{11,11}^{01,01}&\alpha\beta^* R_{11,00}^{01,01}&\pm|\alpha|^2 R_{11,11}^{01,10}&\pm\alpha\beta^* R_{11,00}^{01,10}\\
		\alpha^*\beta R_{00,11}^{01,01}&|\beta|^2 R_{00,00}^{01,01}&\pm\alpha^*\beta R_{00,11}^{01,10}&\pm|\beta|^2 R_{00,00}^{01,10}\\	
		\pm|\alpha|^2 R_{11,11}^{10,01}&\pm\alpha\beta^* R_{11,00}^{10,01}&|\alpha|^2 R_{11,11}^{10,10}&\alpha\beta^* R_{11,00}^{10,10}\\
		\pm\alpha^*\beta R_{00,11}^{10,01}&\pm|\beta|^2 R_{00,00}^{10,01}&\alpha^*\beta R_{00,11}^{10,10}&|\beta|^2 R_{00,00}^{10,10}
	\end{array}
	\right).
	\ee

{\bf The second outcome $\ket{\Phi_\pm}$.} The correct unitaries are given by:
\be 
U_{\Psi_+\Phi_+}=U_{\Psi_-\Phi_-}=\sigma_x,\  U_{\Psi_+\Phi_-}=U_{\Psi_-\Phi_+}=i\sigma_y,
\ee
which leads to the following state:
\be
\label{b10}
\rho_{A'E}^{\Psi\Phi}(\tau)=\left(
\begin{array}{cc}
	|\alpha|^2 R_{11,11}^{10,10}(\tau)&\alpha\beta^* R_{11,00}^{10,01}(\tau)\\[4pt]
	\alpha^*\beta R_{00,11}^{01,10}(\tau)&|\beta|^2 R_{00,00}^{01,01}(\tau)
\end{array}\right),
\ee
and the coherence of the qubit $A'$ is:
\be
C'_{\Phi}(\tau) = \tr_E[w_{00}^\dagger(\tau)w_{01}^\dagger(\tau)w_{10}(\tau) w_{11}(\tau)R(0)].
\ee

{\bf The second outcome $\ket{\Psi_\pm}$.} The corrective unitaries are now given by:
\be 
U_{\Psi_+\Psi_+}=U_{\Psi_-\Psi_-}=\bm 1,\ U_{\Psi_+\Psi_-}=U_{\Psi_-\Psi_+}=\sigma_z,
\ee
and lead to the following state:
\be
\label{b8}
\rho_{A'E}^{\Psi\Psi}(\tau)=\left(
\begin{array}{cc}
	|\alpha|^2 R_{11,11}^{01,01}&\alpha\beta^* R_{11,00}^{01,10}\\
	\alpha^*\beta R_{00,11}^{10,01}&|\beta|^2 R_{00,00}^{10,10}
\end{array}\right),
\ee
so the coherence is given by: 
\be
C'_{\Psi}(\tau) = \tr_E[ w_{00}^\dagger(\tau)w_{10}^\dagger(\tau)  w_{01}(\tau) w_{11}(\tau)R(0)].
\ee

\section{$\Psi_\pm$ initial state}\label{Psi}


In case the first teleportation process is performed using the Bell state $|\Psi_{\pm}\rangle_{BC}$
(instead of $|\Phi_{+}\rangle_{BC}$ in eq.~(1)), the difference in the whole
scenario boils down to exchanging the conditional evolution operators of the environment
according to:
\be
\hat{w}_{00}(\tau)\leftrightarrow\hat{w}_{01}(\tau), \ \hat{w}_{11}(\tau)\leftrightarrow\hat{w}_{10}(\tau).
\ee
This means that the operators governing the initial decoherence process are now
$\hat{w}_{01}(\tau)$ and $\hat{w}_{10}(\tau)$, influencing the state in eq.~\eqref{stanCE}
accordingly, so that the coherence of the qubit $C$ is now given by 
\be\label{cprim} 
c(\tau)\equiv \tr_E [w_{10}^\dagger(\tau) w_{01}(\tau) R(0)]
\ee
instead of (8).

As before, the amount of coherence present in the state of qubit $A$ at the end of the protocol 
depends on the measurement
outcome in the first teleportation, which determines the decoherence which happens
before the second teleportation, and on the outcome of the last measurement.
If the first measurement outcome is $|\Psi_{\pm}\rangle_{AB}$ then
instead of \eqref{C1} and \eqref{C2} we obtain:
\begin{align}
C_{\Psi}(\tau)&\equiv\tr_E[ w_{01}^{\dagger 2}(\tau) w_{10}^2(\tau) R(0)], \label{C1Psi}\\
 C_{\Phi}(\tau)& \equiv \tr_E[ w_{01}^\dagger(\tau) w_{01}^\dagger(\tau) w_{10}(\tau) w_{01}(\tau)R(0)],\label{C2Psi}
\end{align}
where the indexes $\Phi$ and $\Psi$ indicate two possible second measurement outcomes.
The purifing condition (13) now reads:
\be
[w_{01}, w_{10}]=0,
\ee
leading to 
\be
C_{\Phi}(\tau)=1.\label{PT'}
\ee

If the first measurement outcome does not match the initial Bell state
then the coherence of qubit $A'$ at the end is given by 
\begin{align}
C'_{\Psi}(\tau)&\equiv\tr_E[ w_{10}^\dagger(\tau) w_{00}^\dagger(\tau) w_{11}(\tau) w_{01}(\tau)R(0)], \label{C1'Psi}\\
 C'_{\Phi}(\tau)& \equiv \tr_E[ w_{10}^\dagger(\tau) w_{11}^\dagger(\tau) w_{00}(\tau) w_{01}(\tau)R(0)].\label{C2'Psi}
\end{align}
As before, each of the cases \eqref{C1Psi}, \eqref{C2Psi}, \eqref{C1'Psi}, \eqref{C2'Psi} happen with the same probability of $1/4$ so that the purifying teleportation \eqref{PT'} happens in $25\%$ cases.

\section{$C$-factors in the spin-boson model}\label{Cfactors}

Decoehernce factors \eqref{C1}, \eqref{C2} and \eqref{C1'}, \eqref{C2'} correspond to a ``stroboscopic decoherence'' -- a decoherence process, interrupted by measurements. They
contain four conditional evolution operators $w_{ij}$ unlike the usually studied decoherence factors which contain two. We will find them for the $\Phi_\pm$ initial state as studied in the main text. The calculations for $\Psi_\pm$ are analogous. 
Let us first recall the well known form of $w_{ij}$. We will first slightly change the notation for the later convenience,
orginizing the double index $ij$ into a vector $\bm\epsilon =(\epsilon_1,\epsilon_2)$, where $\epsilon_m=\mp 1/2$ corresponds to $i=0,1$. In the interaction picture,  $w_{\bm\epsilon}$ are then given by \cite{reina02} (we are dropping the environmental mode index $k$ for clarity):
\be
w_{\bm\epsilon}^I=D\left[\alpha(\tau) \bm\epsilon\cdot\bm g\right] e^{i\xi(\tau)|\bm\epsilon\cdot\bm g|^2},
\ee
where $D(\cdot)$ is the displacement operator, $\bm g=g( e^{-i k r_1}, e^{-i k r_2})$ is the vector of coupling constants, and the time-dependent functions read:
\begin{align}
\alpha(\tau)&=\frac{1-e^{i\omega \tau}}{\omega},\\
\xi(\tau)&=\frac{\omega \tau - \sin\omega \tau}{\omega^2}.
\end{align}
We calculate
\be
\tr[w_{\bm a}^\dagger w_{\bm b}^\dagger w_{\bm c} w_{\bm d} R(0)].
\ee
Since $R(0)$ is a thermal state, it commutes with the free Hamiltonian of the environment and we can use the interaction picture. Using multiple times the composition law for the displacement operators $D(a)D(b)=e^{(ab^*-a^*b)/2} D(a+b)$ 
and the thermal average formula:
\be
\tr[D(a)\varrho_T]=\exp\left[-\frac{|a|^2}{2} \coth(\frac{\beta\omega}{2})\right]
\ee
we find:
\begin{align}
\tr[w_{\bm a}^\dagger w_{\bm b}^\dagger w_{\bm c} w_{\bm d} R(0)]&=e^{i(\phi_0+\phi+\phi'+\phi_+)}\times\nonumber\\
\exp\left[-\frac{|\alpha(\tau)|^2}{2}\left| (\bm c + \bm d - (\bm a + \bm b))\cdot \bm g\right|^2\coth(\frac{\beta\omega}{2})\right],\label{wstring}
\end{align}
where
\begin{align}
\phi_0&=\xi(\tau)\left(|\bm c \cdot \bm g|^2 + |\bm d \cdot \bm g|^2-|\bm a \cdot \bm g|^2-|\bm b \cdot \bm g|^2\right)\\
i\phi&=\frac{|\alpha(\tau)|^2}{2}\left[(\bm a \cdot \bm g)(\bm b \cdot \bm{g^*}) - (\bm a \cdot \bm{g^*})(\bm b \cdot \bm{g})\right]\\
i\phi'&=\frac{|\alpha(\tau)|^2}{2}\left[(\bm c \cdot \bm g)(\bm d \cdot \bm{g^*}) - (\bm c \cdot \bm{g^*})(\bm d \cdot \bm{g})\right]\\
i\phi_+&=\frac{|\alpha(\tau)|^2}{2}\big[((\bm a +\bm b)\cdot \bm {g^*})( (\bm c +\bm d)\cdot \bm{g}) -
((\bm a +\bm b)\cdot \bm{g})( (\bm c +\bm d)\cdot \bm{g^*})\big].
\end{align}
Calculation is now straightforward. For example, $C_\Phi$ corresponds to $\bm a =\bm b = (\frac{1}{2},\frac{1}{2})$, $\bm c = \bm d = (-\frac{1}{2},-\frac{1}{2})$. The phase factors in \eqref{wstring} all vanish and
\begin{align}
&&|\alpha(\tau)|^2&=\frac{2(1-\cos\omega\tau)}{\omega^2},\\
&& \left| (\bm c + \bm d - (\bm a + \bm b))\cdot \bm g\right|^2 &= 8|g|^2[1+\cos(k (r_1 - r_2))].
\end{align}
Restoring the multimode character of the environment, labeled by the index $k$, introducing spectral density $J(\omega)=\sum_k |g_k|^2 \delta(\omega-\omega_k)$, and the time-of-flight parameter $\bar t$ via $k|r_1-r_2|=\omega \bar t$, we obtain from \eqref{wstring}:
\be
\ln C_\Phi=
-8 \int d\omega J(\omega) \frac{1-\cos\omega\tau}{\omega^2}\coth(\frac{\omega\beta}{2})\left(1+\cos\omega\bar t\right).
\ee
The other factors are calculated analogously. For all of them the phase factors in \eqref{wstring} vanish and the resulting quantities are real. 
In particular,  $C_{\Psi}(\tau) \equiv \tr_E[ w_{11}^\dagger(\tau) w_{00}^\dagger(\tau) w_{11}(\tau) w_{00}(\tau)R(0)]$ corresponds to
$\bm a =\bm c = (\frac{1}{2},\frac{1}{2})$, $\bm b = \bm d = (-\frac{1}{2},-\frac{1}{2})$ for which
\be
\left| (\bm c + \bm d - (\bm a + \bm b))\cdot \bm g\right|^2=0
\ee
so that $C_{\Psi}(\tau) =1$. This looks formally like a Decoherence Free Subspace (DFS) condition but it is defined in a two-step decoherence process.

\section{The general $d$-dit case}\label{ddit}

Here we study a generalization to arbitrary $d$-dimensional systems, using the original teleportation protocol from \cite{bennett93}. It is defined by the following measurements and correction unitaries:
\begin{align}
\ket{\Psi_{nm}}&=\frac{1}{\sqrt d}\sum_{j=0}^{d-1} e^{\frac{2\pi ijn}{d}} \ket{j,j\oplus m}, \label{Pnm}\\
U_{nm}&=\sum_{j=0}^{d-1} e^{\frac{2\pi ijn}{d}} \ket j \bra{j\oplus m}. \label{Unm}
\end{align}
where $n,m\in [0,d-1]$ just like the basis indicies $j,j',k,k',\cdots$ and $\oplus$ is the addition modulo $d$ here.
We start with the following initial state:
\be \label{stan0d}
\sigma(0)=\ket{\psi}_A\bra\psi\otimes\ket{\Psi_{00}}_{BC}\bra{\Psi_{00}}\otimes R(0),
\ee
where 
\be
\ket\psi=\sum_{j=0}^{d-1}\psi_j\ket j
\ee
and $\ket{\Psi_{00}}$ denotes now the maximally entangled state $1/\sqrt d \sum_j \ket{jj}$, according to the notation of \eqref{Pnm}, which we will use throughout this Section. From \eqref{U}, \eqref{Pnm}, and \eqref{Unm}) it is easy to see that the total state after the first teleportaion step and conditioned on the measurement result $nm$ is given by:
\be
\tilde\sigma_{nm}(\tau)=
 \ket{\Psi_{nm}}_{AB}\bra{\Psi_{nm}}\otimes \frac{1}{d^2}\sum_{j,j'}\psi_j\psi_{j'}^* \ket j_C\bra{j'}\otimes R_{j\oplus m, j'\oplus m},
\label{snm'}
\ee
where 
\be
R_{ij}=w_{ii}(\tau) R(0)w^\dagger_{jj}(\tau).
\ee
The first step thus introduces the following decoherence factors into the teleported state:
\be 
c^{m}_{jj'}(\tau)=\tr_E\left[ w^\dagger_{j'\oplus m, j'\oplus m} w_{j\oplus m, j\oplus m} R(0)\right],\label{c1}
\ee 
(they do not depend on the index $n$ which is responsible for
differenciating the entangled states by the phase factors).

In the second step, we first prepare the additional qudits $A'B'$ in the state corresponding to the first measurement, i.e. if the measurement result was $nm$, the initial state is
given by $\ket{\Psi_{nm}}_{A'B'}\bra{\Psi_{nm}}\otimes R_{k\oplus m, k'\oplus m}$.  The decoherence process \eqref{U} then affects the entangled state. It is important that it couples to the same  environment $E$ as in the first step.
The resulting state reads:
\begin{align}
 U_{A'B'E}\tilde\sigma_{nm}(\tau)U^\dagger_{A'B'E}&=\frac{1}{d^3} \sum_{jj'kk'} e^{\frac{2\pi in}{d}(k-k')} \psi_j\psi^*_{j'} \ket k_{A'} \bra{k'}\nonumber\\
 &\bm\otimes \ket{k\oplus m, j}_{B'C}\bra{k'\oplus m,j'} \bm\otimes w_{k,k\oplus m}R_{j\oplus m, j'\oplus m}w^\dagger_{k',k'\oplus m}.
\label{snm''}
\end{align}
Now we perfom the second teleportation: i) We measure the system $B'C$ in the basis $\ket{\Psi_{nm}}$; let us assume the result is $n'm'$; ii) We apply a correction unitary from the set \eqref{Unm}. We note that  unlike in the first step, the entangled resource is now a general state $\ket{\Psi_{nm}}$ (cf. \eqref{snm'}) rather then $\Psi_{00}$ to which the sets \eqref{Pnm}, \eqref{Unm} were tailored. Because of this, the corresponding correction unitary is in general not given by $U_{n'm'}$ but is some permuted one $U_{NM}$, where the indicies $N,M$ will be determined below to reproduce the initial state $\ket \psi$. To avoid excessive algebraic complications in one step, we first apply the measurement and the corrective operation to the appropriate systems:
\be
\langle \Psi_{n'm'}|k\oplus m,j\rangle \cdot U_{NM} \ket k=
\frac{1}{\sqrt d}\sum_{lr}e^{\frac{2\pi i}{d}(-n'l+Nr)} \delta_{l,k\oplus m}\delta_{j,l\oplus m'}\delta_{k,r\oplus M} \ket r \label{2st}
\ee
The delta symbols force that $k=r\oplus M$ and $j=r\oplus M \oplus m \oplus m'$. Applying \eqref{2st} to \eqref{snm''}, we obtain the following (unnormalized) state of $A'E$: 
\be
\label{rAEd1}
\tilde\varrho_{A'E}^{(2)}(\tau) =
 \sum_{jj'rr'} \frac{\psi_j\psi^*_{j'}}{d^4}  e^{i\varphi}\delta_{j,r\oplus M \oplus m \oplus m'}
\delta_{j',r'\oplus M \oplus m \oplus m'} \ket r \bra{r'}
\bm\otimes w_{r\oplus M,r\oplus M\oplus m}R_{j\oplus m, j'\oplus m}w^\dagger_{r'\oplus M,r'\oplus M\oplus m},
\ee
where for brevity's sake we neglected the indices $nmn'm'$ labeling the state $\tilde\varrho_{A'E}$ just substituting them with the superscript $\phantom\rho^{(2)}$ indicating the sate after the second step of the protocol. The phase $\varphi$ is equal to:
\be
\varphi=\frac{2\pi }{d}\bigg[n\left(r\oplus M-r'\oplus M\right)
-n'\left(r\oplus M\oplus m - r'\oplus M \oplus m\right)+N(r-r')\bigg]
 \equiv \frac{2\pi}{d} (r-r')\Big[n-n'+N\Big], \label{phi2}
\ee
since the differences of modular sums generate at most an extra $\pm d$ term, which in turn shifts the phase by an irrelevant factor $\pm 2\pi (n-n')$. To complete the teleportation, we have to find the proper values of the indices $N,M$, properly correcting the state of $A'$. From \eqref{phi2} it is clear that in order to get rid of the phase factor, $N$ must be chosen as:
\be
N=n'\ominus n =(n'-n)\text{mod}d.\label{N}
\ee
Looking at the arguments of delta symbols in \eqref{rAEd1}, we see that in order to reproduce the correct state on $A'$, we must have $r=j$ and $r'=j'$. Let us choose 
\be 
M=\left[d-(m\oplus m')\right]\text{mod}\, d. \label{M}
\ee
Then it is easy to check that 
\be
M \oplus m \oplus m'= 0,
\ee
because by definition $a=m\oplus m' \in [0, d-1]$ and $M\oplus a=d-a+a-d =0$  from the definition of the modular sum (the other case is trivial as for $a=0$, $M=0$ from its definition). Hence 
\be 
r\oplus M \oplus m \oplus m' = r
\ee
in \eqref{rAEd1}, forcing through the delta symbols that $r=j$ and $r'=j'$ as needed. We thus finally obtain:
\be
\label{rAEd2}
\tilde\varrho_{A'E}^{(2)}(\tau)=
\frac{1}{d^4} \sum_{jj'} \psi_j\psi^*_{j'}  \ket j \bra{j'}
\bm\otimes w_{j\oplus M,j\oplus M\oplus m}R_{j\oplus m, j'\oplus m}w^\dagger_{j'\oplus M,j'\oplus M\oplus m},
\ee
so that the protocol outputs the following family of teleported states:
\be
\varrho^{A'}_{mm'}=\sum_{jj'} \psi_j\psi^*_{j'}  C^{mm'}_{jj'} \ket j \bra{j'}, \label{rhofinal}
\ee
where:
\be
C^{mm'}_{jj'}(\tau)=
\tr_E\left[w_{j\oplus M,j\oplus M\oplus m}w_{j\oplus m,j\oplus m }R(0) w^\dagger_{j'\oplus m, j'\oplus m}w^\dagger_{j'\oplus M,j'\oplus M\oplus m}\right].\label{bigC}
\ee
are decoherence factors between the levels $jj'$ and $M$ is defined by \eqref{M}.  We note that the teleported states do not depend on the indicies $n,n'$ so there is a $d^2$-fold degeneracy, leading to $d^2/d^4=d^2$ distinct states with equal probabilities $1/d^2$

Let us study the possibility of reversing the decoherence. Analogously to the qubit case, we assume the commutativity condition
\be 
[w_{ii}(\tau),w_{jj}(\tau)]=0, \text{ for every }i,j \label{commut'}
\ee
motivated by the fact that it is satisfied in many important models. Then, the only possibility to reverse the decoherence completely, i.e. on all the off diagonal terms of \eqref{rhofinal}, is to find combinations of $m,m'$, such that the corresponding states have all of the $d(d-1)/2$ decoherence factors of the form :
\be  
\tr_E \left[ w_{ll} R_{kl} w_{kk}^\dagger \right]=\tr_E \left[w_{kk}^\dagger w_{ll}^\dagger w_{kk} w_{ll} R(0)\right]. \label{C=1}
\ee
Looking at \eqref{bigC}, the necessary condition for this is 
\be 
m=0
\ee 
as $m$ is shifting the second index of $w_{j\oplus M,j\oplus M\oplus m}$ and we need both indicies to be the same. 
For $m=0$ the decoherence factors take the form:
\be 
\tr_E\left[w^\dagger_{j'j'} w^\dagger_{j'\oplus M,j'\oplus M}w_{j\oplus M,j\oplus M}w_{jj} R(0)\right].\label{m=0}
\ee
In order for all of them to be of the form \eqref{C=1}, the following condition would have to be satisfied:  
\be 
\forall jj', j\ne j'\  \exists M\ne 0 : j=j'\oplus M \text{ and } j'=j\oplus M \label{warunek}
\ee
($M=0$ leads to the factors containing $[w^\dagger_{kk}]^2w_{ll}^2\ne \bm 1$, cf. \eqref{C1}). In particular, it would imply that:
\be 
M=j\ominus j'
\ee
and on the same time $M$ would be independent of $jj'$ for all $j\ne j'$. But this can happen only in $d=2$, where there are only two levels and one nonzero $M=1$ switching between them. 

One may ask if a partial purification is possible, when only some of the coherences are restored. This is indeed the case as one of the two conditions \eqref{warunek} can always be satisfied for any $M$. It happens for the levels $jj'$ of the form:
\be
(j, j\oplus M), 0\leq j \leq d-1. \label{1step}
\ee
For every such a pair $jj'$, the decoherence factor \eqref{m=0} does not in general reduce to unity like it was in the qubit case but instead reduces under \eqref{commut'} to the one-step form \eqref{C1}:
\be 
\tr_E\left[ w^\dagger_{j\oplus 2M,j\oplus 2M} w_{jj} R(0)\right].\label{c1'}
\ee 
There are in general $d$ such reduced decoherence factors in each of the $d$ states \eqref{rhofinal}  with $m=0$, $m'$ arbitrary. In particular, for a qutrit $d=3$ all decoherence factors are reduced in each of the $3$ states  $(m=0, m')$  as there are $d(d-1)/2=3$ off-diagonal elements in each state.

For an even dimension $d$, a further, full reduction can happen for  a special state with $M=d/2$. Indeed, since $M=d/2$, $2M=d$ and $j\oplus M=j$ in \eqref{c1'}. Due to the symmetry around $d/2$, the distinct  levels for which this happens are given by:
\be 
(j,j+d/2), 0\leq j < d/2, \label{jj'}
\ee
and there are $d/2$ of them in the state $m=0, M=d/2=m'$ (cf. \eqref{M}). The rest of  the $d(d-2)/2$ decoherence factors are not reduced. We recall that we were considering above only cases with $m=0$ as otherwise there is no chance for even a partial purification under \eqref{commut'}. 

Summarizing, for an arbitrary dimension $d>2$  a partial supression/reversal of decoherence is possible.  It can happen only for those instances when the first Bell measurement gives $m=0$ result. We then have:
\begin{itemize}
	\item For every $m'$, there will be  levels $jj'$ in the corresponding state $\varrho_{0m'}$ \eqref{rhofinal}, such that the decoherence factors \eqref{bigC} between them reduce to the one-step form \eqref{c1'}, i.e. like if only one decoherence process has happened. There are $d$ such levels, described by \eqref{1step}.
	
	\item Additionally, when $d$ is even, the state corresponding to $m'=d/2$ will have $d/2$ of its coherences  \eqref{jj'} fully restored to the original form like if no decoherence has happened at all.
\end{itemize}

\end{document}